\documentclass[aps,prb,twocolumn,showpacs,showkeys,groupedaddress]{revtex4}

\usepackage{graphicx}% Include figure files
\usepackage{dcolumn}% Align table columns on decimal point
\usepackage{bm}% bold math
\usepackage[center,small]{subfigure}
\usepackage{epsfig}
\usepackage{float}
\usepackage[vflt]{floatflt}
\usepackage{longtable}%

\usepackage{multirow}

\begin{document}

\title {Double superconducting transition in the filled skutterudite PrOs$_4$Sb$_{12}$ and
sample characterizations}

\author{M.-A. M\'{e}asson}
\email{marie_audemeasson[at]yahoo.fr}
\affiliation{D\'{e}partement de Recherche Fondamentale sur la Mati\`{e}re Condens\'{e}e, SPSMS, CEA Grenoble, 38054 Grenoble, France}
\affiliation{Graduate School of Science, Osaka University, Toyonaka, Osaka, 560-0043, Japan}
\author{D. Braithwaite}
\affiliation{D\'{e}partement de Recherche Fondamentale sur la Mati\`{e}re Condens\'{e}e, SPSMS, CEA Grenoble, 38054 Grenoble, France}
\author{G. Lapertot}
\author{J.-P.~Brison}
\affiliation{D\'{e}partement de Recherche Fondamentale sur la Mati\`{e}re Condens\'{e}e, SPSMS, CEA Grenoble, 38054 Grenoble, France}
\author{P. Bordet}
\affiliation{Institut NEEL, CNRS-UJF, BP166, 38042 Grenoble cedex 9, France}
\author{H.~ Sugawara}
\affiliation{Faculty of Integrated Arts and Sciences, The University of Tokushima, Tokushima 770-8502, Japan}
\author{P. C. Canfield}
\affiliation{Ames Laboratory and Department of Physics and Astronomy, Iowa State University, Ames, Iowa 50011, USA}
\author{J. Flouquet}
\affiliation{D\'{e}partement de Recherche Fondamentale sur la Mati\`{e}re Condens\'{e}e, SPSMS, CEA Grenoble, 38054 Grenoble, France}

%\date{\today}

\begin{abstract}

A thorough characterization of many samples of the filled skutterudite compound PrOs$_4$Sb$_{12}$ is provided.
We find that the double superconducting transition in the specific heat ($T_{c1}\sim1.89~$K and $T_{c2}\sim1.72~$K) tends to appear in  samples with
a large residual resistivity ratio, large specific heat jump at the superconducting transition and with the highest absolute value of the specific heat above $T_{c1}$. However, we present evidence which casts doubt on the intrinsic nature of the double
superconducting transition. The ratio of the two
specific heat jumps $\frac{\Delta C(T_{c1})}{\Delta C(T_{c2})}$ shows a wide range of values on crystals from
different batches but also within the same batch. This ratio was strongly reduced by polishing a sample down to 120$\mu$m. Remarkably, three samples exhibit a single sharp transition of $\sim$15~mK in width at $T_{c}\sim$~1.7~K. The normalized specific heat jump $\left.C-C_{normal}/C_{normal}\right)_{T_c}$ of two of them is higher than $\sim32\%$ so larger than the sum of the two specific heat jumps when a double transition exists. As an evidence of better quality,
the slope in the transition is at least two time steeper.
%The double superconducting transition may be due to two segregated parts of the sample with distinct $T_c$.
 We discuss the origins of the double transition; in particular we consider, based on X-ray diffraction results, a scenario involving Pr-vacancies. The superconducting phase diagram under magnetic field of a sample with a single transition is fitted with a two-band model taking into account the good values for the gap as deduced from thermal conductivity measurements.

%Thanks to this finding, we have drawn precisely the first superconducting hase
%diagram under magnetic field of a single transition sample.

\end{abstract}

\pacs{65.40.Ba%65.40.Ba Heat capacity
%,74.25.Fy %Transport properties (electric and thermal conductivity, thermoelectric effects, etc.)
%,74.25.Ha %Magnetic properties
,71.27.+a%strongly correlated electron system; heay fermions
,74.25.Dw%superconducting phase diagram
%,74.25.Op%Mixed states, critical fields, and surface sheaths
,74.70.Tx%heavy femrions superocnductor
,74.20.Rp% Pairing symmetries (other than s-wave)
%74.62.Fj% Pressure effects
}
\keywords{PrOs$_4$Sb$_{12}$, skutterudite, unconventional superconductivity, specific heat}

\maketitle

\section{Introduction}

Since the discovery of the first Pr-based heavy fermion superconductor
PrOs$_{4}$Sb$_{12}$ (T$_{c}\sim 1.85$K) by Bauer et al. \cite{Bauer2002}, this system has attracted much attention
with particular emphasis on the possible unconventional nature of superconductivity.
 A significant piece of the evidence for unconventional superconductivity is the double superconducting transition
 seen in specific heat first reported in 2003 by Vollmer et al. \cite{Vollmer2003} and Maple et al. \cite{Maple2002}. Ever since, a plethora of publications have dealt with its observation and with possible theories.
This double transition has since been observed by specific heat measurements by many
groups (from Japan~\cite{Aokiprivatediscussion}, USA \cite{Maple2002, Rotundu2006}, Germany \cite{Grube2006, Cichorek2005, Vollmer2003,Drobnik2005}, France \cite{Measson2004}) and by thermal expansion~\cite{Oeschler2004} with samples from different origins and even in La doped or Ru substituted samples \cite{Frederick2005condmat,Rotundu2006}.
So whatever the origin, the double transition is a robust property of this compound and we will see that it appears in good samples.

However susceptibility measurements on a sample with a very
clear double superconducting transition~\cite{Measson2004} induced the first doubts about
its microscopic origin. Indeed even in this good sample, the diamagnetism is not perfect at $T_{c1}$, the highest superconducting transition temperature, and a second step at $T_{c2}$, the lowest one, is
visible. This caused us to look more closely at other evidence for a microscopic origin, including the magnetic field, and pressure dependences of the double transition. The phase diagram under magnetic field didn't bring evidence for its intrinsic nature since the field dependence of $T_{c2}$ is completely similar to that of $T_{c1}$. The shape of $H_{c2}(T)$ can be quantitatively understood with a two-band model~\cite{Measson2004} and indeed the same model simply scaled with $T_c$ has been used to fit both lines.
The behavior of $T_{c1}-T_{c2}$ under pressure is also not conclusive \cite{MeassonICM}. At low pressure, the slope $\partial
T_{c1}/\partial P$ is at least 20\% smaller than $\partial
T_{c2}/\partial P$. However, above 1~GPa the behavior of the two transitions is similar, with
$T_{c1}-T_{c2}$ stabilizing around 200~mK. These results do not rule out an intrinsic origin but certainly provide no supporting evidence towards it, (contrary to the well documented case of UPt$_3$ where the different field and pressure dependences of the two transitions were decisive results).

We report here on a study of the nature of the double
superconducting transition of PrOs$_4$Sb$_{12}$. Our main purpose is
to clarify whether the double superconducting transition which
appears in specific heat is intrinsic, like it is now admitted for
UPt$_3$ or extrinsic, due to sample inhomogeneities, as shown for
URu$_2$Si$_2$~\cite{Ramirez} and high-$T_c$ superconductor YBa$_2$Cu$_3$O$_x$~\cite{Janod1993}. We discuss this point from systematic
characterizations by resistivity, specific heat and susceptibility
measurements. From general characterizations, we conclude that the
double transition appears in good samples. But an extensive study
particularly of many small samples provides evidence which brings
strong doubts about its microscopic origin; the most convincing one is
the existence of three samples with a single sharp superconducting
transition with a $T_c$ matching $T_{c2}$. Then we present the single
crystal x-ray diffraction results. Finally we show the
first measurement of the phase diagram under magnetic field for a
sample with a single sharp superconducting transition, and a fit of
the upper critical field with a two band model.

%According to our characterizations, the quality of these samples is better that of the samples showing the double transition:
%the slope of %the specific heat in the transition is at least twice steeper than the steepest transition in the double transition samples
% and the %specific heat jump normalized to its value in normal state is larger than the sum of the two jumps in the double
%transition samples. Moreover, we were able to dramatically change the ratio of the two specific heat jumps by polishing a small
%sample. All our observations are consistent with an extrinsic double transition, $T_{c2}$ being the intrinsic transition
%temperature. Its origin might be the presence Pr-vacancies in the structure, a feature frequently encountered in filled
%skutterudites compounds. We present here a single crystal x-ray diffraction study in order to discuss this statement.

\begin{figure}[!h]
\begin{center}
\includegraphics[width=0.5\textwidth]{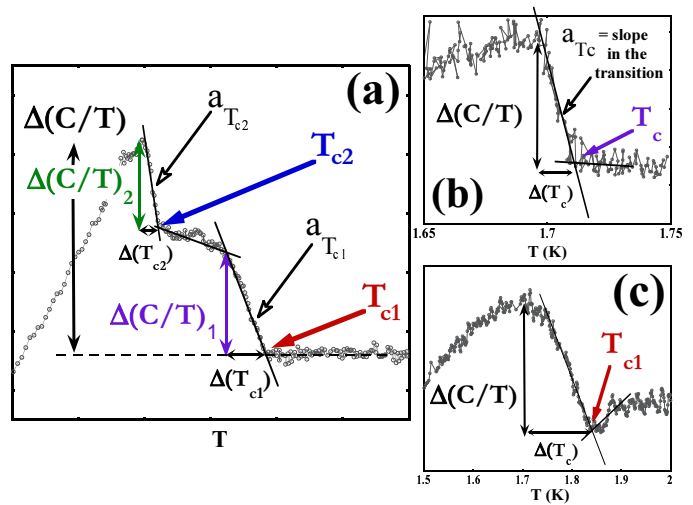}
\end{center}
\caption{(color online) Definition of the criteria we use in this paper: (a) for the samples with a double superconducting transition, (b) with a sharp single transition and (c) with a broad single transition.}
\label{notationtotale}
\end{figure}

We would like first to introduce some criteria we will use in this
paper. Figure~\ref{notationtotale} provides the main criteria
depending on the kind of superconducting transition (double, single
and broad, or single and sharp), i.e. the superconducting transition
$T_c$, $T_{c1}$ and $T_{c2}$ obtained on the onset, the specific
heat jumps $\Delta(C/T)_1$ at $T_{c1}$, $\Delta(C/T)_2$ at $T_{c2}$
and $\Delta(C/T)$ for the whole jump, the width of the transition
$\Delta(T_{c1})$ at $T_{c1}$, $\Delta(T_{c2})$ at $T_{c2}$ or
$\Delta(T_{c})$ when a single transition appears, and the slope in
the transition $a$. This last criterion appears more relevant than
the simple width of the transition when we compare one of the jumps
in the double transition with a single sharp jump, which is probably two times
higher. The residual resistivity ratio $RRR$ is always measured between 300~K and 2~K.
%To get the normalized specific heat, we substracted the normal part obtained under magnetic field up to 1~T. As noticed
%in~\cite{Grube2006}, the shape of the specific heat in the normal part is not affected under this magnetic field.

%dans cette publi

\section{\label{sct:carac}General Characterizations}

\begin{figure}[!h]
\begin{center}
\includegraphics[width=0.5\textwidth]{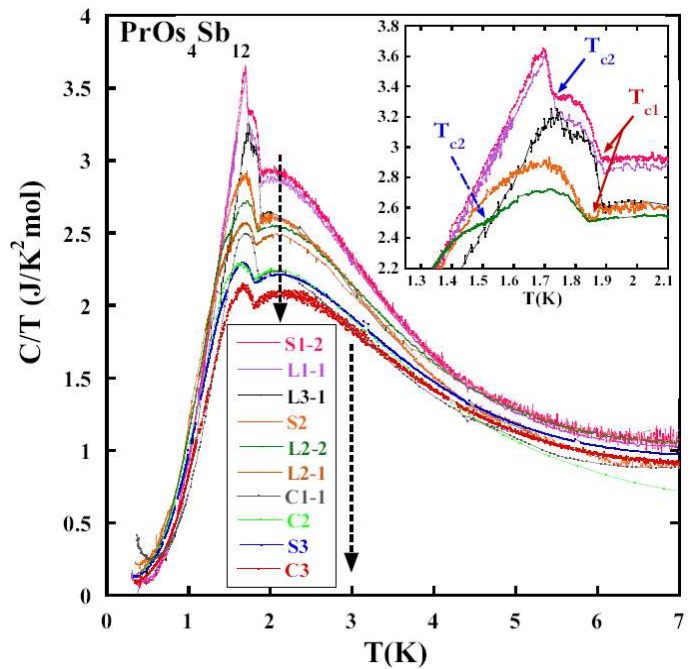}
\end{center}
\caption{(color online) Specific heat as $C/T$ versus $T$ for several samples of PrOs$_4$Sb$_{12}$ from different batches. The samples' labels follow the decreasing order of the specific heat at 2~K. The absolute value of $C$ clearly depends on the sample, with a value on the Schottky anomaly at 2~K included between 2.06~J/K$^2$.mol and
2.94~J/K$^2$.mol. The insert is a zoom on the superconducting jump for the samples with the highest specific heat. A double superconducting transition appears in some samples while the other samples exhibit a broad single transition.}
\label{Cabsolues}
\end{figure}

The crystals of PrOs$_4$Sb$_{12}$ were grown by the Sb flux method
\cite{Sugawara2002,Takeda1999,Bauer2001} by 3 separate groups (P.
Canfield, H. Sugawara and G. Lapertot, respectively labelled C, S and
L). The first number indicates the batch. When there were 2
different samples from the same batch we add an extra index.
The crystals from P. Canfield and G. Lapertot were separated from flux by a hot spinning process. Remaining flux droplets were dissolved in hydrochloric/water
solution. The samples used in ref.~\cite{Measson2004}, \cite{Seyfarth2007}, \cite{Huxley2004},
\cite{MeassonICM} are respectively S1, L1-1A, S3, L1-6.

The specific heat ($C$)
measurements were performed in a $^{3}He$ calorimeter either by a quasi-adiabatic method with a
Au/Fe-Au thermocouple controlled by a superconducting quantum
interference device (SQUID) or by a heat-pulse relaxation technique using a Physical Property Measurement System
(PPMS) from Quantum Design.

Figure \ref{Cabsolues} presents the specific heat of several samples
of PrOs$_4$Sb$_{12}$. Their
absolute values at 2~K differ strongly, varying from
2.06~J/K$^2$.mol to 2.94~J/K$^2$.mol and can vary on the
samples in the same batch (see samples L2-1 and L2-2). Other
published results report absolute values at 2~K between
1.3~J/K$^2$.mol \cite{Maple2002} and 3.2~J/K$^2$.mol
\cite{Grube2006}. This discrepancy cannot be explained only by the
presence of trapped Sb flux. The specific heat broad peak at 2~K is mainly due to the Schottky term due to the presence at
$\sim~8K$ of the first crystalline electric field (CEF) excited
level of the 4f$^2$ Pr states, $\Gamma_{4}^{(2)}$, above the singlet
ground state $\Gamma_1$. As we point out to possible
Pr-vacancies in the structure (see section~III.B), we propose that
the samples dependence of the specific heat is partly due to these
Pr-vacancies. Quantizing the necessary percentage of vacancies to fit
the curves is a difficult task. Indeed, some distortions may appear
with Pr-vacancies and they may change locally the CEF, which broaden the Schottky term. Moreover the interactions between Pr ions are quite strong as shown by the dispersion of the inelatic spectrum in wavevectors~\cite{Kuwahara2005}; so the consequence of the Pr-vacancies cannot be reproduced by a pure local model. Roughly a maximum of 10~\% of Pr-vacancies is
required to fit the specific heat of sample L3-1 to L1-1. No upturn
due to the nuclear Schottky anomaly except a small feature for the
samples of batch C1, was observed, at least above 0.4 K. We also
note that no anomaly was detected at 0.6~K, temperature at
which several experiments report a change of behavior \cite{Cichorek2005}.  %discuter plus ça?

The double superconducting transitions for several samples are shown in the zoom of Fig.~\ref{Cabsolues} (see insert). The shape of the double transition of samples from different origins (L1-1 and S1-2) is similar but is quite different from sample L3-1 with similar $T_c$s and even larger deviations are found from sample L2-2 both in absolute value of $C$ and in value
of $T_{c2}$. The sample dependence of the shape of the double transition will be discussed in detail in section~\ref{sct:DT}.

\begin{figure}[!h]
\begin{center}
\includegraphics[width=0.45\textwidth]{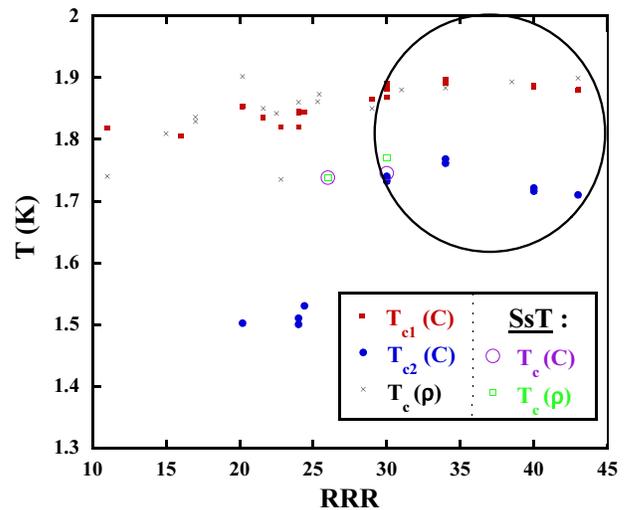}
\caption{Superconducting transition temperature versus the RRR between 300~K and 2~K determined by specific heat (full square and circle) or resistivity (cross) measurements. The open circles and squares show the $T_c$s determined by $C$ and $\rho$ measurements of the samples with a single sharp superconducting transition L1-5 and L1-1A. Definition of the $T_{c}$s is shown in the Fig~\ref{notationtotale}. The large circle includes all the samples with a clear double superconducting transition (excluding batch L2) and sample L1-1A. The general tendency is that the double transition appears in the samples with the highest RRR.}
\label{TcversusRRR}
\end{center}
\end{figure}

\begin{figure}[!h]
\begin{center}
\subfigure[\label{Cschottky}]
{\includegraphics[width=0.45\textwidth]{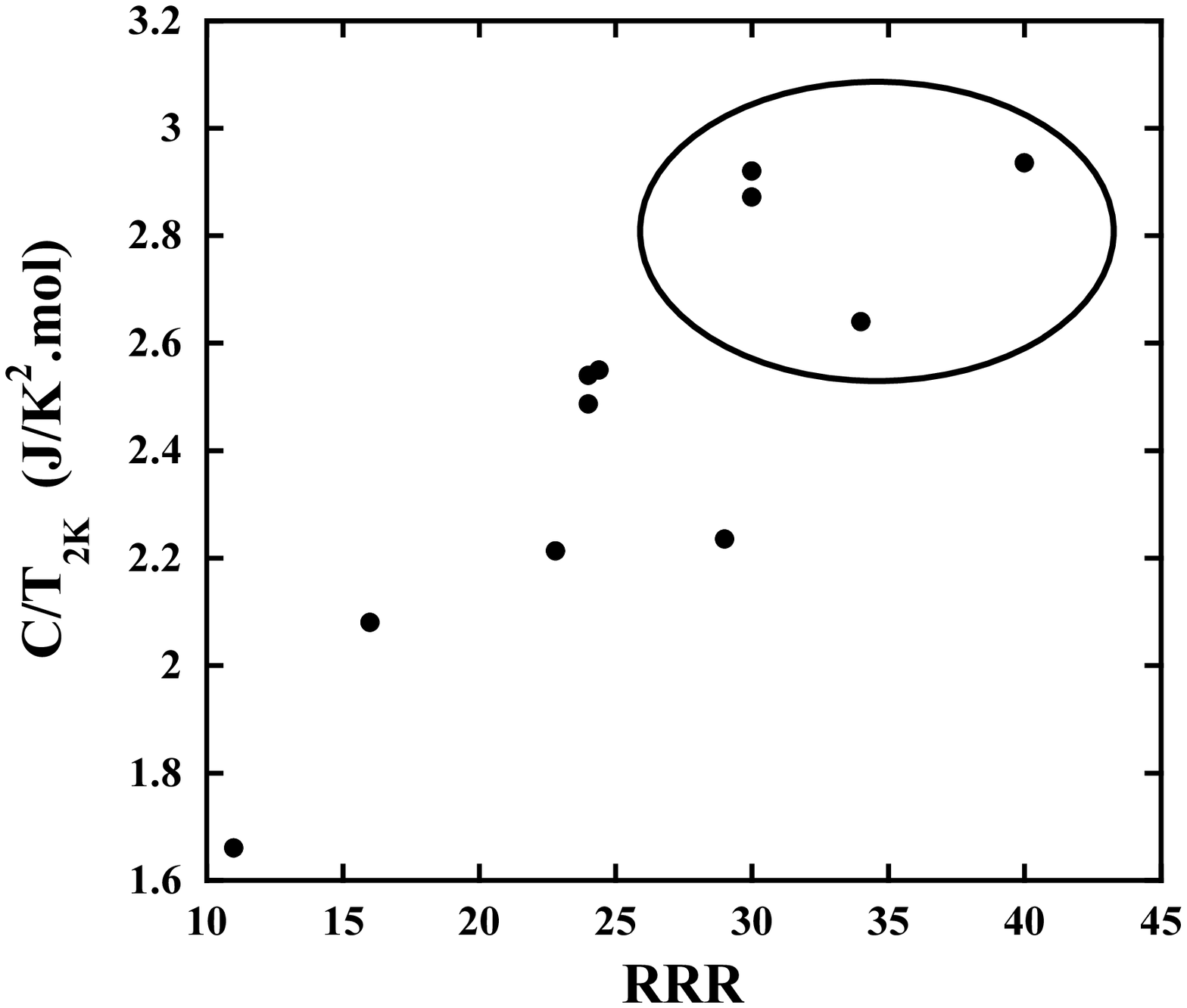}}
\subfigure[\label{Cjump}]
{\includegraphics[width=0.45\textwidth]{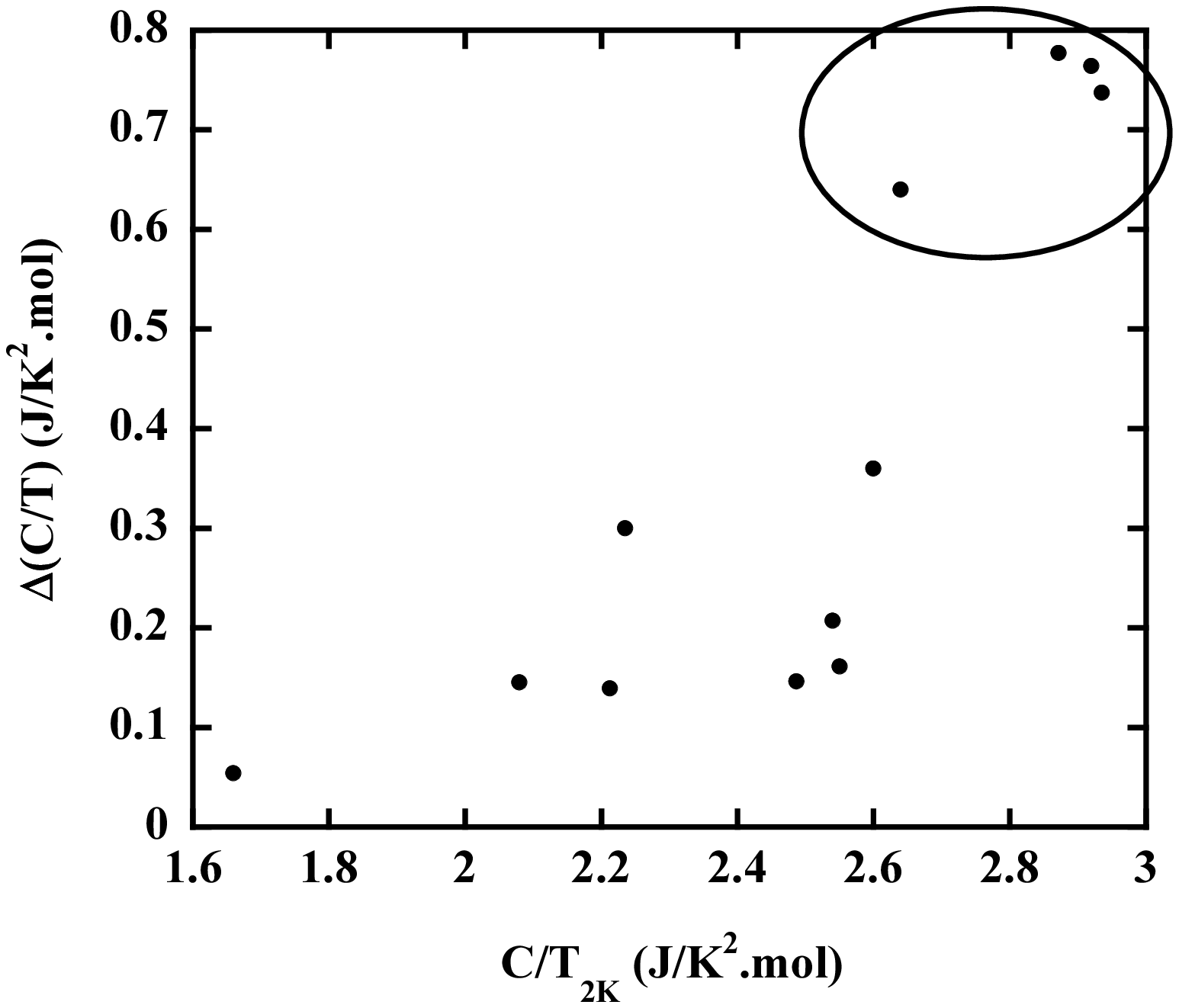}}
\caption{(a) Absolute value of the specific heat $C/T$ on the Schottky anomaly at 2~K versus the $RRR$. The highest the $RRR$ is, the larger the specific heat at 2~K is. (b) Entire specific heat jump $\Delta(C/T)$ (defined figure~\ref{notationtotale}) versus the specific heat $C/T$ at 2~K. The two large circles embody the samples with a clear double superconducting transition (excluding batch L2) : the samples with a double transition are characterized by the highest $C$ at 2~K and the highest specific heat jump at the superconducting transition.}\end{center}
\end{figure}
We discuss now the link between the appearance of the double
transition and sample quality. Figure~\ref{TcversusRRR} shows the
superconducting transition temperatures determined by specific heat
and resistivity measurements versus the residual resistivity ratio.
In both cases $T_c$ was obtained from the onset criterion. When the
specific heat and the resistivity were not measured on the same
sample (which was often the case), the $RRR$ values taken are an
average for the batch. As the spread of the $RRR$ is large (for
instance the $RRR$ of batch L1 is included between 17 and 33.8),
only a general tendency can be extracted. However it is quite clear
that $T_{c1}$ increases slightly with the $RRR$ (its minimum and
maximum values are respectively 1.805~K and 1.897~K) whereas
$T_{c2}$ is strongly sample dependent. All the smaller $T_{c2}$ at
about 1.5~K were reported from samples belonging to batch L2. The
large circle indicates all the samples with a clear double
transition (excluding sample L2-3) and also contains a sample with a
single sharp transition L1-1A. The double transition seems to appear
in the samples with the largest $RRR$. We
notice that the samples with a single sharp transition have a $T_c$
similar to $T_{c2}$ and a relatively high but not the highest RRR.
They will be discussed further in section~III.

Figure~\ref{Cschottky} shows the absolute value of $C$ on the Schottky anomaly at 2~K against the $RRR$. The highest the $RRR$ is, the largest the specific heat at 2~K is. Figure~\ref{Cjump} presents the total jump of the specific heat at the superconducting transition ($\Delta(C/T)$), which is probably the best criterion of the quality of the samples, against the value of $C$ at 2~K. Again, the highest the specific heat jump is, the largest the specific heat at 2~K is. Moreover as indicated by the large circles on Fig.~\ref{Cschottky} and \ref{Cjump} which embodies all the samples with a double transition, the double transition seems to be a feature of the samples which meet all these criteria. Samples with a broad single superconducting transition, which {\it always} covers the temperature range of the double transition, are clearly of worse quality than the double transition samples.

However, we will present in the next section some characterisations which provide several pieces of evidence against an intrinsic origin of the double transition, specially those carried out on very small samples (with a typical size of 100~$\mu m$ and lighter than 0.1~mg).
%%%%%

\section{\label{sct:DT}Double superconducting transition}

\subsection{Characterization results}

The first doubts about the intrinsic nature of the double
superconducting transition came from the results of susceptibility
measurements. Indeed no published result shows perfect diamagnetism
at $T_{c1}$ : the ac susceptibility always exhibits a very broad
superconducting transition or a double step matching with $T_{c1}$
and $T_{c2}$ of the specific heat results
\cite{Measson2004,Cichorek2005}.  All the samples with a double
transition we have tested by ac susceptibility ($\chi$), i.e.
batches L1, L2, S1, also exhibit a double step matching the specific
heat jumps. We report here other stronger doubts, particularly based
on the existence of samples with a single sharp superconducting
transition. We will compare their quality with those of samples with
a double transition.

%general view on the table:
Table~\ref{tab} provides a general view and the main features of all
the samples with a double transition or a single sharp
superconducting transition we have tested. All of them except S1-1A
and L1-1A are as grown (unpolished) samples. All the samples not
reported in table~\ref{tab} exhibit a single transition with a width
larger than the temperature range of the double transition as shown
in Fig.~\ref{Cabsolues}. The criteria for $T_{ci}$,
$\Delta(T_{ci})$, $\Delta(C/T)_i$, $a_{T_{ci}}$ are defined in the
Fig.~\ref{notationtotale}.

We first notice that the double transition appears in 5 different
batches from 3 origins with similar shape for the biggest samples of
the batches L1, S1 and C1. But further investigations, specially on
small samples (with a typical size of 100$\mu$m), show a wide range
of the ratio $\Delta(C/T)_1/\Delta(C/T)_2$ from 0.2 to 3.36. It even
differs in the same batch, from 0.2 to 0.7 or from 0.25 to 3.36 in,
respectively, batches L1 and L2. If we suppose that one of the
superconducting transitions is more sensitive to the quality of the
sample than the other, other properties should also be affected.
However it clearly appears that we cannot connect the value of the
ratio with other criteria, as the value of $T_{c2}$ or as the width
of the transition at $T_{c2}$, $\Delta_{T_{c2}}$, by comparing,
respectively samples L2-2 and L2-3 or samples L3-1, L1-1 and L1-6.

\newbox\hautbox \setbox\hautbox=\hbox{\vphantom{\rule[-.15cm]{0cm}{.5cm}}} %pour créer une boite de largeur nulle qu'on insère ensuite dans les lignes
%du tableau: permet de regler la hauteur des lignes
%\newbox\hautbox1 \setbox\hautbox=\hbox{\vphantom{\rule[-0.2cm]{0cm}{0.2cm}}}
\newbox\htbox \setbox\htbox=\hbox{\vphantom{\rule[-.1cm]{0cm}{0.4cm}}}
\newbox\hatbox \setbox\hatbox=\hbox{\vphantom{\rule[-.2cm]{0cm}{0.5cm}}}

\begin{table*}%[h] add [H] placement to break table across pages
\begin{ruledtabular}
\begin{tabular}{| @{\usebox{\htbox}}l|l|c|c|c|c|c|c|c|c|}

\multicolumn{1}{| @{\usebox{\hatbox}} l|}{~sample}
&\multicolumn{1}{ @{\usebox{\hatbox}} c|}{\tiny{shape~\textvisiblespace~size/weight}~\textvisiblespace~\tiny{measurement}}
&\multicolumn{1}{ @{\usebox{\hatbox}} c|}{~$T_{c1}${\tiny(K)}}
&\multicolumn{1}{ @{\usebox{\hatbox}} c|}{~$T_{c2}${\tiny(K)}}
&\multicolumn{1}{ @{\usebox{\hatbox}} c|}{~$\frac{\Delta(C/T)_1}{\Delta(C/T)_2}$}
&\multicolumn{1}{ @{\usebox{\hatbox}} c|}{~$\Delta(T_{c1})${\tiny(mK)}}
&\multicolumn{1}{ @{\usebox{\hatbox}} c|}{~$\Delta(T_{c2})${\tiny(mK)}}
&\multicolumn{1}{ @{\usebox{\hatbox}} c|}{~$\frac{\Delta (C/T)}{C/T_{normal}}$ }
&\multicolumn{1}{ @{\usebox{\hatbox}} c|}{~$a_{T_{c2}}${\tiny~or} $a_{T_c}${\tiny(K$^{-1}$)}}&{$a_{T_{c1}}${\tiny(K$^{-1}$)}} \\
\hline
\multicolumn{10}{ @{\usebox{\hatbox}} c}{Double Superconducting Transition} \\
\hline
{ S1-2} &{AofC.~\textvisiblespace~1.97mg}&{1.884} &{1.721} &{1.23} &{61} &{23}&{30.6$\%$}&{4.1}&{1.98}\\%02-002 2cubes %%fait

{ L1-1} &{AofC.~\textvisiblespace~$\sim$10mg}&{1.890} &{1.737} &{0.70} &{61} &{35}&{32.3$\%$}&{4.23}&{1.58}\\%lot "L1" aggrégat de cubes %%ùfait
{ L1-3} &{c.~\textvisiblespace~200$\mu$m}&{1.868} &{1.74} &{0.29} &{43} &{40}&{30.7$\%$}&{4.7}&{1.3}\\%lot "L1" petit cube %%fait
{ L1-4} &{b.~\textvisiblespace~500$\mu$m}&{1.889} &{1.732} &{0.20} &{64} &{35}&{28$\%$}&{5.2}&{0.79}\\%lot 1 gde barrette%%%fait
{ L1-6}& {p.~\textvisiblespace~50*150*150$\mu$m$^3$~\textvisiblespace~AC}&{1.85} &{1.73}&{0.54} &{36} &{39}&{$\geq$27.2$\%$}&{3.7}&{2.0} \\%%%fait

%{"L1-5"} &{ $\emptyset$ } &{1.732} &{\bf{0}} &{ $\emptyset$ } &{0.017}&{>26$\%$}&{-2.6}&{$\emptyset$} \\%lot "L1" plaquette ; magik fromCp12P %%%fait
{ L3-1} &{AofC.~\textvisiblespace~1.25mg}&{1.891} &{1.761} &{2.8} &{46} &{21}&{27.3$\%$}&{3.7}&{3.4}\\%lot "L3" aggrégat de cubes PPMs %%%%fait (rapport des sauts changés de 3.6 à 2.8) précision sur pente 2 mauvaise
{ L3-2} &{b.~\textvisiblespace~200$\mu$m}&{1.897} &{1.76} &{1} &{60} &{40}&{32$\%$}&{4.1}&{2.14}\\%lot "L3" petite barrette %%%fait
\hline

\multicolumn{10}{@{\usebox{\hautbox}}c}{   Sharp Single Superconducting Transition} \\
\hline
{ L1-5} &{p.~\textvisiblespace~{160*200*40$\mu$m$^3$}~\textvisiblespace~PPMS} &{$\emptyset$}&{1.733} &{\bf{0}} &{ $\emptyset$ } &{\bf{17}}&{33$\%$$\leq$ $\leq$44$\%$}&{\bf{16$\pm$4}}&         {$\emptyset$} \\%lot "L1" plaquette ; magik PPMS en haut, ac en bas %%%fait
{}&{ \textvisiblespace~AC}&{} &{}&{} &{} &{}&{$\geq$28$\%$}&{\bf{16.5$\pm$2}}&{}\\%mettre les chiffres entres le 2 lignes

{ L1-7}& {p. \textvisiblespace ~ 50*150*150$\mu$m$^3$  \textvisiblespace~  AC}&{$\emptyset$} &{1.680}&{\bf{0}} &{$\emptyset$} &{21}&{$\geq$22$\%$}&{\bf{8.1}}&{$\emptyset$} \\%lot "L1 éclat saut pris sur courbe corrigé %%%fait

{ L1-1A}&{p.\textvisiblespace~45*150*200$\mu$m$^3$/$\sim$0.1mg}&{1.800} &{1.745} &{{\bf$\sim$0} ($\leq$0.09)} &{55} &{27}&{36$\%$}&{\bf{10$\pm$2}}&{$\emptyset$}\\%lot ponce par JP brison dans sa publi de kappa %%fait
\hline

\multicolumn{10}{@{\usebox{\hautbox}}c}{   Double Superconducting Transition with a weaker $T_{c2}$} \\
\hline
{ L2-2}&{b.~\textvisiblespace~9.7mg}&{1.844} &{1.53} &{3.36} &{150} &{105}&{$16\%$}&{0.3}&{1.05}\\%lot "L2-2 barre %%%fait
{ L2-3}&{b.~\textvisiblespace~150$\mu$m}&{1.85} &{1.535} &{0.25} &{75} &{47}&{$28\%$}&{4.6}&{1.75}\\%lot "L2-3 barrette%%%fait

{ C1-2}&{c.~\textvisiblespace~200$\mu$m}&{1.877} &{1.685} &{1.17} &{95} &{35}&{$27\%$}&{2.2}&{1.4}\\%lot "C1 cube %%%fait
\hline
\multicolumn{10}{@{\usebox{\hautbox}}c}{   Effect of polishing : change of the relative height of the two specific heat jumps} \\
\hline
{ S1-1} &{p.\textvisiblespace~300$\mu$m~\textvisiblespace~AC}&{$\sim$1.88} &{$\sim$1.71} &{{\bf1.9}} &{100} &{25}&{$\geq$28$\%$}&{4.2}&{1.5}\\%%normalization mauvaise et ac; attention au %TC %%% fait pente prise sur normalizé rapport des sauts pris sur mesure brute
{ S1-1A} &{p.~\textvisiblespace~120$\mu$m~\textvisiblespace~AC}&{$\sim$1.88} &{$\sim$1.71} &{{\bf$\sim$1.0}} &{180$\pm$20} &{20}&{$\geq$28.1\%}&{7.74}&{1.0$\pm$0.3}\\%%attention au Tc! %%%fait pente prise sur normalizé rapport des sauts pris sur mesure brute

\end{tabular}
\caption{Main features of the specific heat of the samples with a double superconducting transition or a single sharp transition.
To get S1-1A and L1-1A, S1-1 and one piece of L1-1 were respectively polished.
Abbreviations are: c.=cube; b.=bar, AofC.=aggregate of cubes, p.= platelet; AC=ac calorimetry measurements.
The slope at $T_{ci}$, $a_{T_{ci}}$ is taken on the specific heat jump normalized to its normal value. The
criteria are indicated in the Fig.~\ref{notationtotale}. \\
A large range of ratio of the two specific heat jumps appears. Three samples with a single sharp transition exist: L1-5, L1-7 and L1-1A. The superconducting transition of sample L1-5 is
the sharpest. For samples L1-5 and L1-1A and relatively to the specific heat in the normal state, the specific heat jump is higher than the sum of the two specific heat jump in the double transition samples. The slope in the superconducting transition at $T_{c2}$, $a_{T_{c2}}$, increases
with decreasing ratio $\Delta(C/T)_1/\Delta(C/T)_2$, reaching the highest values for the 3 samples with a single transition (the transition  is even three times steeper for the single transition sample L1-5).}
\label{tab}
\end{ruledtabular}
\end{table*}%l'étoile change tout, elle impose que ce soir étaler sur les 2 colonnes.
%The $RRR=\rho_{300K}/\rho_{2K}$ is 43 for S1-1, 26 for L1-5, 30 for L1-1A and 24.4 for L2-2. The absolute value of the specific
%heat $C/T$ at 2~K is 2.93~J/K$^2$.mol for S1-2, 2.87~J/K$^2$.mol for L1-1, 2.54~J/K$^2$.mol for L2-2, 2.64~J/K$^2$.mol for L3-1.

\begin{figure}[!h]
\begin{center}
\includegraphics[width=0.5\textwidth]{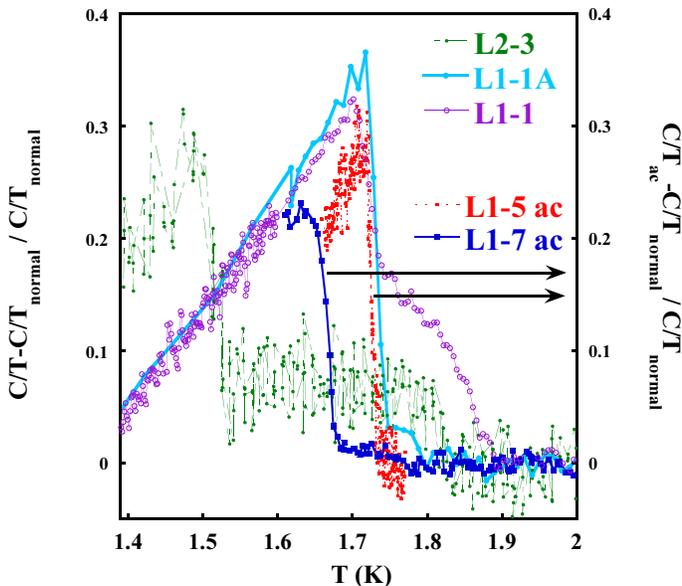}
\caption{(color online) Zoom on the superconducting transition of the specific heat normalized to its normal value of selected samples of PrOs$_4$Sb$_{12}$. Samples L1-1 and L2-3 exhibit a double transition, with the most common shape for L1-1 and with a quite different $T_{c2}$ for L2-3. Our remarkable finding is the existence of three samples L1-5, L1-7 and L1-1A with a single sharp transition (with a tiny remaining jump at 1.8~K for L1-1A). It clearly questions the intrinsic nature of the double transition. Two single transition samples L1-5 and L1-7 were measured by an ac method. Their specific heat jumps are underestimated. The specific heat of sample L1-5 was also determined semi-quantitatively by relaxation method (Cf. figure~\ref{4circles}). All features are reported in table~\ref{tab} and comparison of the quality of the samples is provided in the text.}
\label{DTnorma}
\end{center}
\end{figure}

\begin{figure}[!h]
\begin{center}
%\subfigure[\label{montageX}]{\includegraphics[width=0.2\textwidth]{montX.eps}}
%\subfigure[\label{montagerho}]{\includegraphics[width=0.2\textwidth]{rho.eps}}\\
\includegraphics[width=0.5\textwidth]{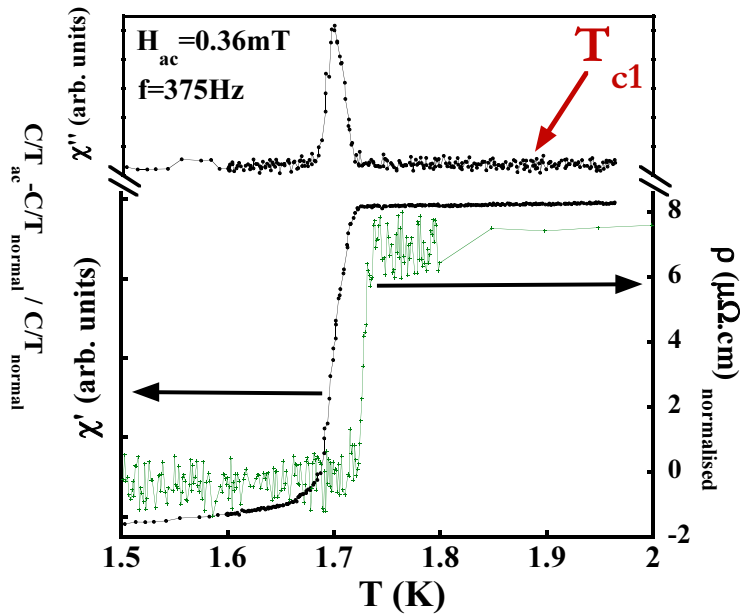}
\caption{Ac susceptibility and resistivity of the sample L1-5. The width of the transition in resistivity is 16~mK and 35~mK in susceptibility.
$T_c$ is consistent for all measurements ($C$, $\rho$ and $\chi$). The RRR is 26. There is no trace of superconductivity  above $T_c$. This sample has a single superconducting transition which happens to be the sharpest in all measurements.}
\label{xrhomagik}
\end{center}
\end{figure}

%focus on single transition sample
Before discussing further the characterizations, we focus on our most remarkable finding which is shown in Fig.~\ref{DTnorma}: we measured the specific heat of three samples L1-5, L1-7 and L1-1A with a single sharp
superconducting transition (L1-1A has still a tiny jump at 1.80K as well as a small step in resistivity at 1.85K~\cite{Seyfarth2007}). The specific heat of a sample with a "usual" double transition (L1-1) is also presented. We notice the existence of sample L2-3 with a clear double transition but with a quite reduced $T_{c2}$. L1-5 and L1-7 are very small as-grown platelets with well-developed faces and with a thickness of about 50$\mu$m. L1-1A has been obtained
by polishing a large cube (1~mm) of the sample L1-1 so that
the thickness was reduced down to 45~$\mu m$ \cite{Seyfarth2007}. Their critical temperature $T_c$
is 1.733~K, 1.680~K and 1.745~K, so matching $T_{c2}$ of the samples with a double transition.
One of these samples L1-5 was further characterized by single crystal x-ray diffraction, ac susceptibility and resistivity measurements.
We confirmed the composition of the sample (see section~III.B). %Pierre bordet data?
%From the specific heat results, it was a mean to check the diamagnetism.
 For the susceptibility measurement a tiny susceptometer in glass fiber was built to get a good
filling ratio. The secondary coil has an external diameter of 500~$\mu$m and hole of 300~$\mu$m. 440 turns of 14~$\mu$m diameter copper wires were wound on each part on the secondary coil. The frequency of the exciting magnetic field of 0.36~mT is about 375~Hz. %The phase %of the signal is corrected by 3$^{\circ}$.
Ac susceptibility and resistivity results are visible in Fig.~\ref{xrhomagik}. From these results,
we checked that this sample has a single superconducting
transition. Indeed, the susceptibility exhibits no sign of superconductivity above $T_c$. All $T_c$ ($\rho$, $\chi$, $C$) are consistent.

%\multicolumn{2}{|c|}{Texte sur 2 colonnes} \\

%%come back to the table to compare the single transition sample and to give other evidence for extrinsic DT

Now we compare the quality of these single transition samples with the samples with a double transition. As the quantitative value of the specific heat of most of the samples is indeterminable (because many samples are too small or were measured by ac calorimetry), we have subtracted the normal part obtained under magnetic field up to 1~T then normalized to this value. We are only able to consider semi-quantitative values because, as we discussed in section~II, the absolute value
of the specific heat depends strongly on the sample.
%To account for the quality of the samples, we used, amongst other criteria, the steepness of the superconducting transition as the slope
 %, %$a_{T_{ci}}$ in figure~\ref{notationtotale}, which is taken on the specific heat normalized to its normal value and in the
%jump of the i$^{th}$ transition.
 The ac calorimetry results of samples L1-7 and L1-6 provide only a bottom limit of the normalized specific heat jump as the background was not subtracted. As for the sample L1-5, a semi-quantitative measurement of $C$ (its mass, about 0.1~mg, was too small to be measured precisely) was performed with the PPMS (Cf. figure~\ref{4circles}). The most important feature is certainly
that the normalized specific heat jump, $\Delta C/C$)$_{T_c}$, is at least as high as the entire transition of the double transition samples. Indeed even if the smallness of the sample L1-5 implies a noisy specific heat measurements by the PPMS (shown in Fig~\ref{4circles}),
the specific heat jump at $T_c$ was found to be 33-44$\%$ of the normal state specific heat, a value larger than the entire jump, $\Delta C/C$)$_{T_c}$, in any sample showing a double transition (maximum of $\Delta C/C$)$_{T_c}$=32.2$\%$).
%the normalized specific heat jump of L1-5 appears
%to be included between $33\%$ and $44\%$, so superior than the highest of the double transition samples (=32.2$\%$).
The conclusion for L1-1A is similar with a $\Delta C/C$)$_{T_c}$ of 36$\%$. The specific heat of L1-7 was only measured by ac method without subtracting the background; the specific heat jump is underestimated.

%\begin{figure}[!h]
%\begin{center}
%\includegraphics[width=0.2\textwidth]{poterybis.eps}
%\caption{Small home made susceptometer in glass fibers.
%Secondary and primary coils are respectively on the left and on the right.}
%\label{montageX}
%\end{center}
%\end{figure}

%compare the width of the transition

As for the width of the transition, sample L1-5 has the sharpest ever measured (16~mK in $\rho$, 17~mK in $C$ and 35~mK in $\chi$).
One remarkable feature is that the transition is three times steeper ($a_{T_c}$ three times larger) than the steepest transition in the samples with a double transition. The values obtained for $a$ may be
influenced to some extent by the range of values of $C$ in the normal
state, but insufficiently to change this semi-quantitative conclusion. Actually, this tendency is confirmed for all
the samples of table~\ref{tab}: the ratio $\Delta(C/T)_1/\Delta(C/T)_2$ decreases with increasing steepness of the transition at $T_{c2}$ reaching zero for the three steepest transitions. This criterion as well as the specific heat jump $\Delta C/C$)$_{T_c}$ points to a higher quality of the samples with a single transition.

%which Tc is the intrinsic one
We think that the transition at $T_{c2}$ is the intrinsic one. Indeed the transition at $T_{c1}$ is always broader and less steep than the transition at $T_{c2}$ (Cf. table~\ref{tab}). All the single sharp transition samples have $T_c$
lower than 1.75~K whereas even in the worst samples (see Fig.~\ref{TcversusRRR}) $T_{c1}$ is not smaller than 1.805~K. From this observation, we exclude that the single transition is the transition at $T_{c1}$ shifted to smaller temperature by some impurities effects. It would also be quite surprising that in these single transition samples the transition at $T_{c1}$ simply disappears due to worse quality of the samples; because the single transition should also be broadened and because the thermal conductivity ($\kappa$) measurements clearly point to a higher quality for sample with a single transition called L1-1A ($\kappa/T$$\sim$~70~$\mu$W/K$^2$.cm at 100~mK \cite{Seyfarth2007}) than for a sample which exhibits a large superconducting transition with $T_{c}$ about 1.85~K ($\kappa/T$$\sim$~250~$\mu$W/K$^2$.cm at 100~mK \cite{Seyfarth2005}).

%doubts:
Finally only two points could leave some doubt.
First, the RRR of samples L1-5 and L1-1A are respectively 26 and 30 whereas the largest value we have got (43) is in a double transition sample (S1-1). Of course, the resistivity is not a probe of the whole volume of the sample. For instance, sample S1-1 exhibits a sharp superconducting transition in resistivity ($\rho$=0 at 1.82~K) but two steps in $\chi$  matching with $T_{c1}$ and $T_{c2}$.
%and the 4 wires were contacted on the unpolished part of S1-1 which has a larger specific heat
%jump at $T_{c2}$ or even only a single transition at $T_{c2}$ (Cf. Fig.~\ref{Cpolished}).
Moreover we cannot exclude that another parameter, such as remaining flux or Pr-vacancies, has the opposite effect on the RRR. Secondly, if we assume that the intrinsic $T_c$ is $T_{c2}$, we can compare $T_c$ of the single transition samples with $T_{c2}$ of the double transition samples. It appears that the transition in the single transition samples is shifted to smaller temperature than the transition at $T_{c2}$ of the double transition samples of batch L3 ($T_c$ is smaller than $T_{c2}$ and the transition in batch L3 is not much broader). One can argue that the parts of the sample with different $T_c$ may be affected independently by others impurities not related to the appearance of the double transition. This hypothesis would also explain the existence of such samples as L2-2 and L2-3 with a $T_{c2}$ 15$\%$ smaller than the highest one but with similar $T_{c1}$. Finally, the rareness of the samples with a single sharp transition (to our knowledge, no other observation was reported) is certainly due to their smallness and all the consecutive experimental difficulties.

\begin{figure}[!h]
\begin{center}
\includegraphics[width=0.45\textwidth]{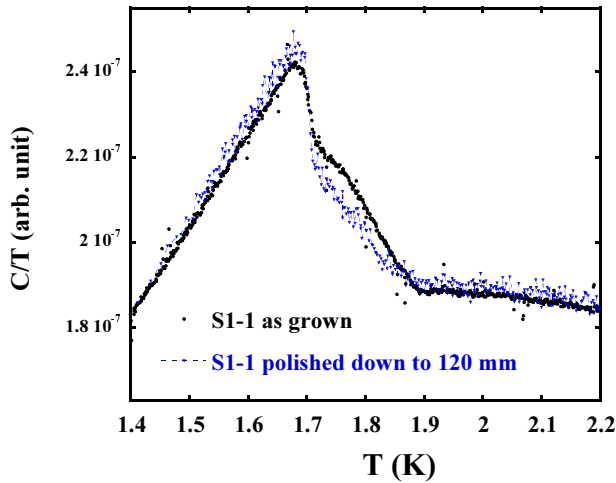}
\caption{AC specific heat versus temperature of samples S1-1 and S1-1A we got after polishing S1-1 down to 120$\mu$m. The ratio of the two specific heat dramatically changes from 1.9 to $\sim$1.0.}
\label{Cpolished}
\end{center}
\end{figure}

Our last proof is presented figure~\ref{Cpolished} showing the ac specific heat of sample S1-1 before and after polishing down to 120$\mu$m (then called S1-1A).
As the samples are too small to determine their mass, the curves are normalized so the entropies in the normal phase match. The data fit in all the temperature
range from 1.3K to 4K except in the double transition. By polishing, a dramatic decrease of the ratio of the specific heat jumps occured, going from 1.9 to $\sim$1.0. The transition at $T_{c2}$ became steeper and sharper whereas we aren't even able to distinguish any change
of slope between $T_{c1}$ and $T_{c2}$ in S1-1A. So polishing the sample S1-1 clearly tended to remove the transition at $T_{c1}$.
%The two phases related with the two different $T_c$ seems to be segregated.

%Finally all these observations are compatible with a double transition due to two parts of the sample
%This was recently confirmed by local magnetization measurements \cite{Kasahara2006}.
Finally all these observations are compatible with a double transition due to two parts of the sample. The effect of polishing presented here as well as reported in~\cite{Seyfarth2007} (we got a single transition sample L1-1A by reducing the thickness down to 50$\mu$m) suggest that the two parts are macroscopically segregated.

We would like again to draw attention to the problem of the quality
of the samples and on possible spurious analyses of the temperature dependence in the superconducting state of the sample with a large single transition. Indeed from the specific heat measurements of batch L2, it appears that some part of the samples can still become superconductor at temperature much lower than $T_{c1}$ (1.52~K for batch L2) and it would not be surprising that it is also the case in the samples with a large single superconducting transition.

%\begin{figure}[!h]
%\begin{center}
%\includegraphics[width=0.5\textwidth]{ratiojumpsverususlope2.eps}
%\caption{}
%\label{}
%\end{center}
%\end{figure}

 %The difference $T_{c1}-T_{c2}$ depends also on the sample. Actually we measured the specific heat of three samples of a second
%batch n$^{\circ}2$. One has a large single transition and two have a double transition with $T_{c1}-T_{c2}$= 315~mK.
%Tc is sample dependent. lot 2: inhomogeneity of batch + ratio of Cp + Tc2 and Tc1-Tc2  sample dependent (NB: susceptibility alos exhibit 2 %steps
%at the 2 transitions)\\
%Moreover we
%found few samples with a $T_{c2}$ 15$\%$ smaller than the usual $T_{c2}\sim 1.7K$.

% Finally the characterization of the polished sample by thermal conductivity and resistivity measurements by %J.-P. Brison et al.
%\cite{Seyfarth2006} gives us another evidence of the higher quality of the samples with a single transition.
% Indeed even if the RRR is only 30, the residual

Of course ruling out the existence of an intrinsic double transition doesn't imply that its superconductivity is conventional.
In particularly, following the idea that $T_{c2}$ is the intrinsic transition, its strong sample dependence (1.76~K for L3-1, 1.71 for S1-1, 1.68 for L1-7 and 1.53~K for L2-2, i.e. a dispersion of 15$\%$) might point to unconventional superconductivity.
Nevertheless this observation must be carefully investigated as the dispersion of $T_{c1}$ is only of 5$\%$ for all our samples. \\

So we have accumulated evidence for an extrinsic double superconducting transition in PrOs$_4$Sb$_{12}$. Thus, in addition to account for the two steps visible in magnetic or resistivty results, the presence of normal part in the sample above $T_{c2}$ would explain the highest temperature minima in flux flow resistance reported in reference~\cite{Kobayashi2005}. %%%hysteresis [11Tayama]% perso je crois pas que %ça explique peak effect, ça explique Ha dans Kobayashi surmeent.

It happens that a definitive conclusion may emerge from a quantitative measurement of the specific heat of a single transition sample, specially by validating our semi-quantitative observation that the single jump at $T_c$ is higher than the sum of the two jumps in the double transition samples. But finding the origin of the extrinsic superconducting transition would provide the definitive answer.

\subsection{\label{subsct:4circles}Origin of the Double superconducting transition : discussion and 4 circles X-rays diffraction study}

As discussed above, the low temperature superconducting transition $T_{c2}$ seems to be the intrinsic one. So a simple random-impurity-induced pair breaking mechanism cannot be responsible for the appearance of the extrinsic transition which occurs at higher temperature ($T_{c1}$) and, whatever the intrinsic transition is at $T_{c1}$ or $T_{c2}$, the narrowness of the lowest-temperature jump rules out this hypothesis, since such an effect would simultaneously broaden and lower $T_c$. So it is not surprising that annealing the sample has no effect on the double transition~\cite{Frederick2005condmat}.%This clearly suggests that each transition is associated with a distinct crystalline %state (macles?).
We note that this case is not isolated in the history of heavy fermions superconductors. The double transition of URu$_2$Si$_2$ was ruled out by Ramirez et al.~\cite{Ramirez} by isolating the lowest $T_c$ phase when removing the surface of the sample. As for CePt$_3$Si, Kim et al. pointed out to a spurious double transition~\cite{Kim2005} due to a second phase of Ce$_3$Pt$_{23}$Si$_{11}$, and $T_c$ decreases from 0.75~K to 0.46~K with increasing quality of the sample~\cite{Takeuchi2007}. Sr$_2$RuO$_4$~\cite{Maeno1996} and CeIrIn$_5$~\cite{Bianchi2001} exhibit a much higher $T_c$ in resistivity measurements than on the specific heat results.
%C' est vrai pour CeIrIn$_5$~\cite{Bianchi2001} ou on observe en resistivité Tc=1K alorsque la chaleur specifique donne 0.4K
%C est la même chose pour Sr2RuO4 3K par rapport à O.5K~\cite{Maeno1996} . On prétend que c est à cause d' inclusion de ruthenium

Multiple scenario are possible like the existence of an impurity phase very similar to PrOs$_4$Sb$_{12}$ and superconductor at $T_{c1}$, or the presence of twins which enhance $T_c$~\cite{AbrikosovTofM}. We test here another hypothetic scenario involving the existence of praseodymium vacancies in the structure, in relation with the disparity of the quantitative value of the specific heat on the Schottky anomaly. Actually in the filled skutterudite structure RT$_4$X$_{12}$, because of a weak interaction of the R-atoms with its neighbours (as points out by the large rattling of the R-atoms in the $X_{12}$ cages), some R-vacancies are commonly observed. Moreover as discussed in section~II, the large dispersion of the specific heat value around 2~K could be ascribed to varying ratio of Pr-vacancies which may strongly affect the Schottky anomaly. Based on these observations, we suggest the overall scenario discussed in\cite{MeassonICM}, namely the transition at $T_{c1}$ appears in the part of the samples with only partial occupancy of the Pr site.

% One possible explanation for these observations might be..., but on further reflection it becomes clear that this is not likely.

\begin{figure}[!h]
\begin{center}
\includegraphics[width=0.5\textwidth]{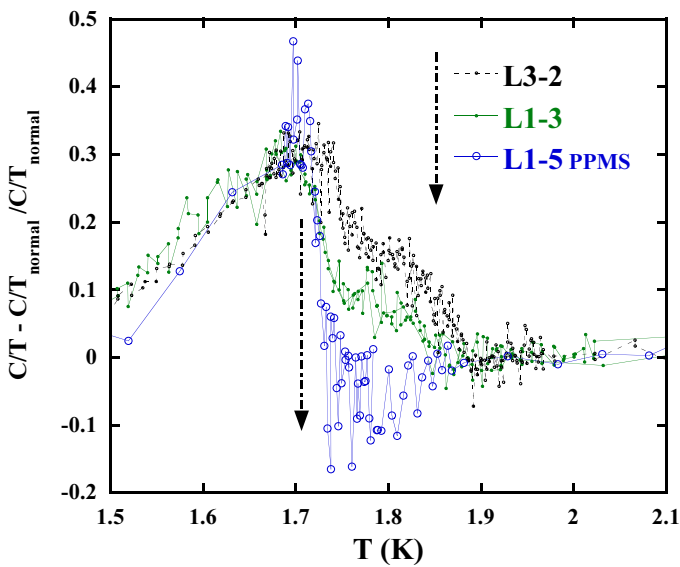}
\end{center}
\caption{Zoom on the superconducting transition in the specific heat normalized to its normal part obtained under magnetic field for the three samples we have selected (for their different ratios of the 2 specific heat jumps) for the 4 circles X-ray diffraction. Results of 4 circles X-ray diffraction are presented table~\ref{Xray}.}
\label{4circles}
\end{figure}

In order to test this hypothesis, we have selected three single crystals with different shapes of the superconducting transition and we have submitted them to a single crystal x-ray diffraction experiment. Fig.~\ref{4circles} shows the superconducting transition in the specific heat normalized to its value in the normal state of these three samples. From sample L3-2 to L1-3, the ratio
$\Delta(C/T)_1/\Delta(C/T)_2$ decreases strongly from 1 to 0.29 reaching zero in sample L1-5. If the hypothesis of Pr vacancies is true, the jump at $T_{c1}$ should increase when the vacancies on the Pr sites becomes larger.

\begin{table}[!h]
\centering
\begin{tabular}{|c|c|c|c|} \hline\hline
 Sample & L3-2    & L1-3 & L1-5  \\\hline
 Cell parameter ($\mathring{A}$) & 9.272(1)    & 9.288(1)& 9.321(1)           \\
occ. (Pr) &       0.93(1)    & 0.89(1) &          0.966(6) \\
U$_{iso}$  (Pr) (${\mathring{A}}^2$) & 0.0370(5)    &0.0359(7) &      0.0384(3)   \\
U$_{iso}$  (Os) (${\mathring{A}}^2$) &0.0046(1)      & 0.00402(5) & 0.00480(3) \\
U$_{iso}$  (Sb) (${\mathring{A}}^2$) & 0.0064(1)  &  0.00578(6)& 0.00660(5)                \\
 x(Sb)($\mathring{A}$)& 0.15608(5)         &  0.15613(3)& 0.15608(2)          \\
 z(Sb)($\mathring{A}$)& 0.34040(5)      & 0.34036(3) & 0.34031(2)            \\
% sin$\theta$ / $\lambda_{max}$  & 1.15&  1.15& 0.97\\
% Ntotal& 20108& 16727 & 8786\\

%Redundancy & 21.3 & 17.9 & 16.2\\
%Rint & 9.3 &  6.3 & 8.03\\
%Nobs & 806 & 860 & 512\\
%Npar & 12 & 12  &12 \\
%Rw & 2.94 & 5.19  & 5.97\\
Gof & 2.24  & 1.95  & 2.33 \\\hline\hline
%sin$\theta$ / $\lambda_{max}$  : maximum value of sin$\theta$ / $\lambda_{max}$.
%Ntotal : total number of reflexions measured.
%Redundancy : redundancy of the collected data set.
%Rint quality factor of the reflexion averaging in point group m-3.
%Nobs : number of observed unique reflexions (I>3.$\sigma(I)$) .
%Npar  : number of parameters in the refinement.
%Rw : weighted R-factor.

\end{tabular}
\caption{ Structural parameters and refinement agreement factors for three crystals. The position of Pr atoms is setting at (0,0,0). U$_{iso}$ : isotropic thermal parameters. The occupancy (occ.) of Os and Sb was set to 1. Gof is the goodness of fit.}
\label{Xray}
\end{table}

The X-ray investigation was carried out with a Nonius KappaCCD
diffractometer equipped with graphite monochromatized AgK$\alpha$ radiation. After sample alignment, up to 20000 Bragg reflexions were collected to
a maximum sin$\theta$/$\lambda$ of 1.15 leading to a very high redundancy. After extraction of the intensities using the EvalCCD software \cite{Duisenberg2003},
a numerical absorption correction was applied using the crystal shape. The structure refinement was carried out using the Jana2000 software \cite{Petricek2000}.
An isotropic extinction correction (type I, Lorentzian distribution) was applied and all atoms were given anisotropic atomic displacement parameters (a.d.p.).
Finally, since an anomalously large a.d.p. was observed for the Pr atom (about 0.04 ${\mathring{A}}^2$) and to test the vacancy ratio on the Pr sites, its occupancy factor was also let to vary.
This systematically led to a slight decrease of the Pr atom occupancy (from 1 to 0.97 for sample L1-5) and the a.d.p. remained practically the
same. The agreement factors were improved, though only slightly : for crystal L1-3 having the lowest refined Pr occupancy (0.89), the goodness of the fit decreased from 2.05 to 1.95 by letting the Pr occupancy parameter free. Table~\ref{Xray} reports the parameters obtained for the three single crystals when the Pr occupancy is refined and the Pr position is set at (0,0,0).

These result confirm that the Pr atom is in a strongly disordered position as reported in~\cite{Cao2003,Grube2006}.
{\it Moreover, they seem to indicate the possibility of a nearly but not completely filled Pr site in the filled skutterudite structure for PrOs$_4$Sb$_{12}$}. In the interests of caution, we should point out that the large rattling of the Pr in the Sb-cages at 300~K may reduce the accuracy of this measurement, and the values may depend slightly on the refinement model. However the occupancy factors reported in table~\ref{Xray} at least support the trend of varying Pr occupancy in these three samples and the Pr occupancy is the highest in sample L1-5 so with a single sharp transition. However a smaller level
of vacancies in sample L3-2 than in sample L1-3 is not relevant with the above-mentioned hypothesis since sample L3-2 has a larger ratio
$\Delta(C/T)_1/\Delta(C/T)_2$. We think that further diffraction measurements at low temperature are required to thoroughly characterize the dynamic and static disorder at the Pr site, and possibly prove or disprove the hypothesis Pr vacancies as the origin of the double transition.

%\cite{Goto2004}: off center posision ??

%By checking various fitting model for samples L1-3 and L3-2, we obtained a qualitative agreement with table ~\ref{Xray}

%Cao: sstatic $\sigma^2$ for PrOs4Sb12 is 0.0045 Å2, lead to an off-center instability for the Pr
%ion.

%%other resutls:

\subsection{Superconducting phase diagram~\label{HT}}

\begin{figure}[!h]
\begin{center}
\subfigure[\label{HTcomparaison}]
{\includegraphics[width=0.4\textwidth]{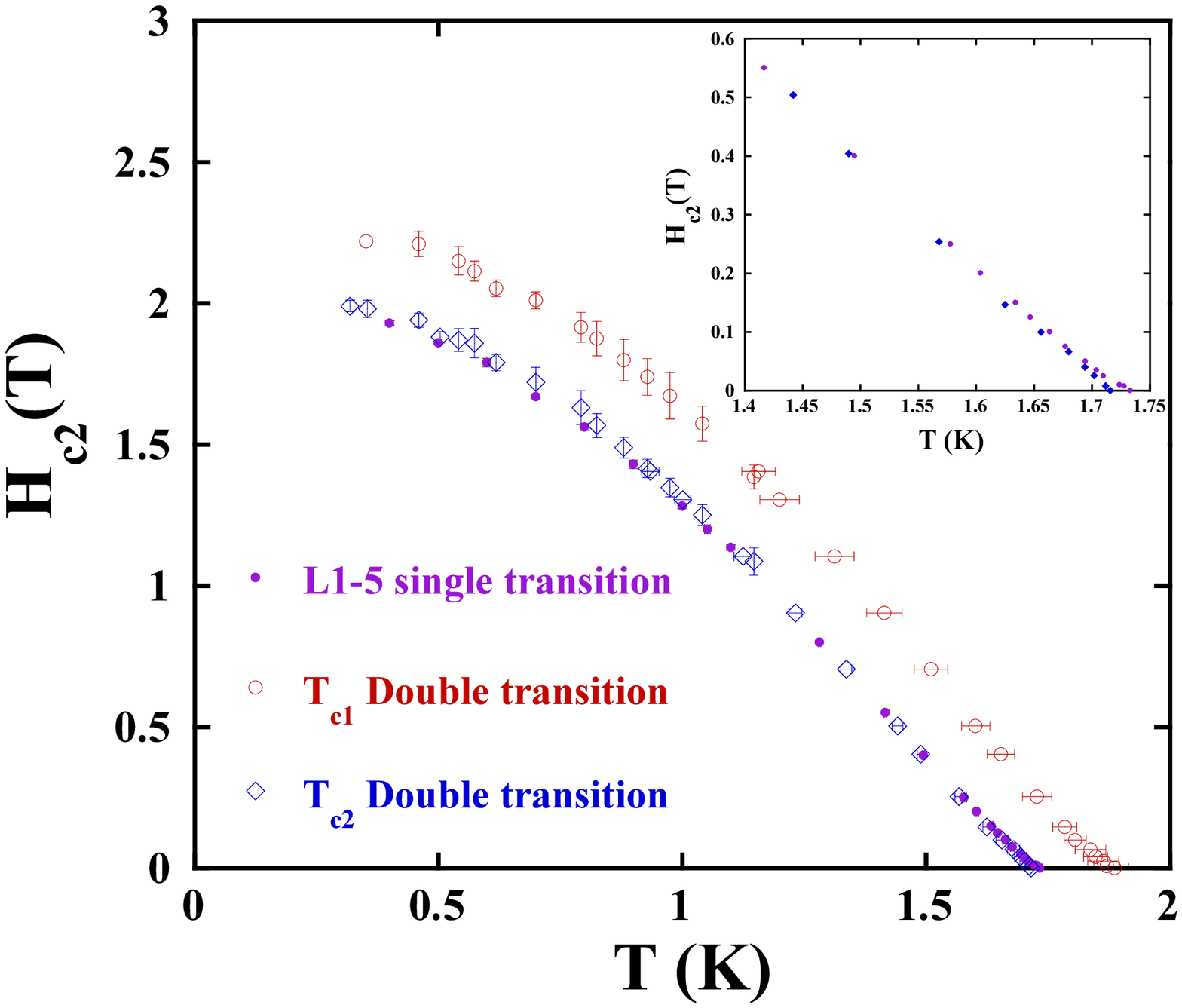}}
\subfigure[\label{HT} ]
{\includegraphics[width=0.4\textwidth]{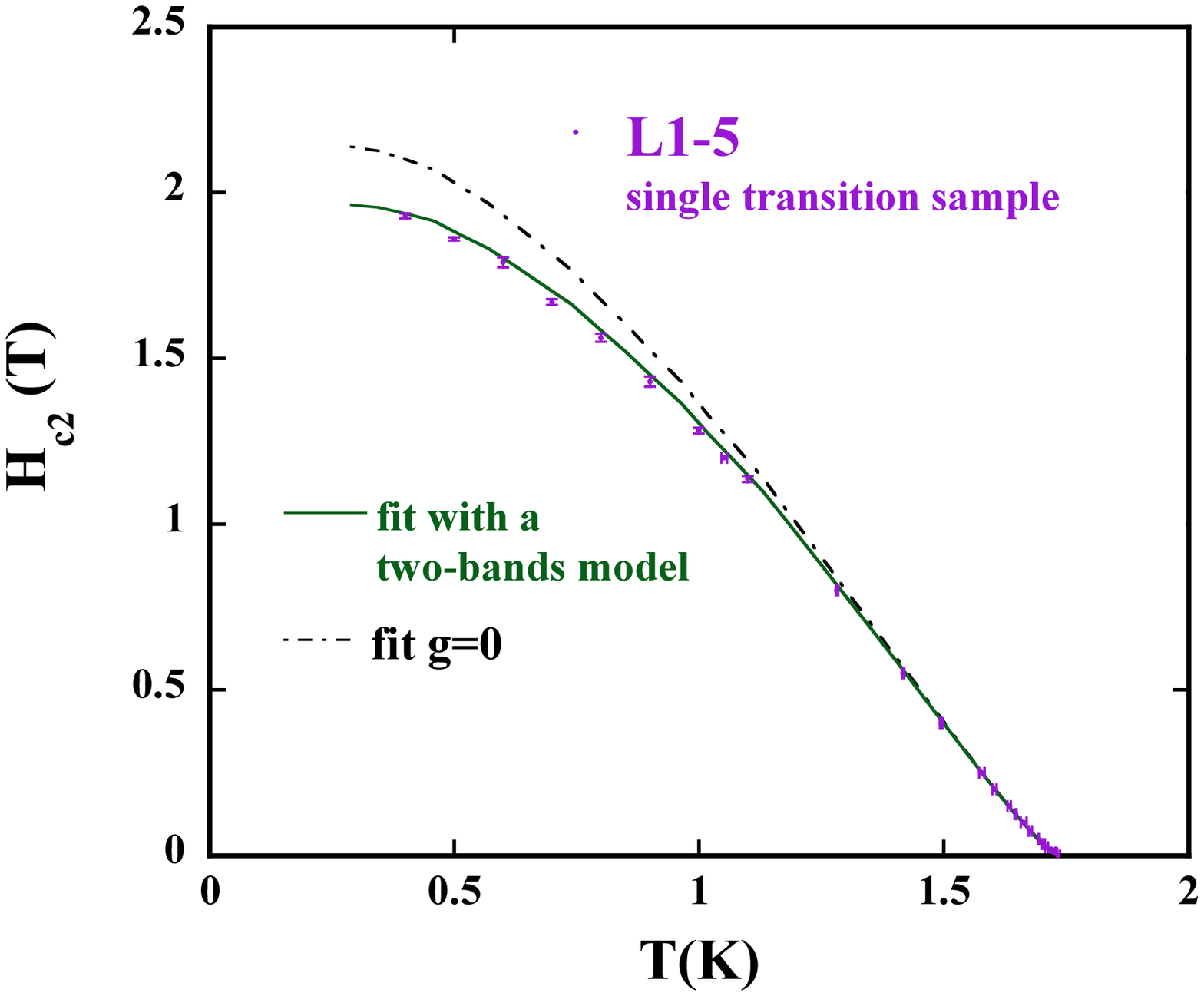}}
\caption{(a) Superconducting phase diagram under magnetic field of the single transition sample L1-5 and the sample with a double transition from~\cite{Measson2004}. The insert presents a zoom on the small positive curvature of both the single transition sample L1-5 and $T_{c2}(H)$ of the sample with a double transition from~\cite{Measson2004}. Clearly the superconducting transition of L1-5 follows the lowest-temperature one of the double transition sample. The small positive curvature at low field is even better resolved. (b) Superconducting phase diagram under magnetic field of a sample with a single sharp transition L1-5 obtained by resistivity measurements. The error barres indicate the witdh of the transition. The data were fitted by a two-band model like in ref~\cite{Measson2004}. The new set of parameters are described in the text. The dashed dotted line shows the fit without Pauli limitation (with g=0)~\cite{note}.}
\end{center}
\end{figure}

We have followed the superconducting transition of the single transition sample L1-5 under magnetic field by resistivity measurements. In figure~\ref{HT}, we report $T_c$ versus $H$, determined by onset criterion. The transition remains very sharp
(12~mK at 1.2~T and less than 30~mT at 400~mK) pointing out again to the high quality of this sample. The $T_c$($H$) line matches with $T_{c2}$($H$) published in~\cite{Measson2004} as shown in figure~\ref{HTcomparaison}. The small positive curvature at low magnetic field is even more clearly visible (Cf. insert of figure~\ref{HTcomparaison}).

This makes it even clearer that multigap effects are disconnected from the question of the double transition. It also backs up the conclusions of the recent thermal conductivity measurements on a high quality-single superconducting transition sample, L1-1A: multigap effects have been confirmed~\cite{Seyfarth2007}, with a very low field scale associated with the light carrier/small gap band, of the same value as found in a previous inhomogeneous (wide specific heat transition) sample~\cite{Seyfarth2005}. Moreover, the fact that the small positive curvature close to $T_c$ is found also in homogeneous samples and with the similar amplitude is a definite proof, beyond the reproducibility of the measurements~\cite{Measson2004}, that it is not connected to sample inhomogeneities.

 It also brings information on the inter band coupling strength different from those of the thermal conductivity experiments. For example, the fit proposed in~\cite{Measson2004} for $H_{c2}(T)$ would also apply to these new measurements, as the data simply scale with $T_c$. But the set of inter and intra band couplings ($\lambda_{ij}$) proposed in this first work, was based on the simplest hypothesis that $\lambda_{ij}$ is proportional to the density of states of band $j$, so that $\lambda_{11} = \lambda_{21}$ and  $\lambda_{12} = \lambda_{22}$. In such a case, a simple calculation of the two gaps in a weak-coupling scheme shows that they are equal. In order to be consistent with the thermal conductivity results, which find a factor three between the small and large gap~\cite{Seyfarth2007}, one needs to introduce a difference between $\lambda_{11}$ and $\lambda_{21}$. The size and position of the curvature on $H_{c2}$ then still impose a very small value of $\lambda_{12}$ (we take still for simplicity ($\lambda_{12} = \lambda_{22}$).

 So, instead of the set of parameters : $\lambda_{11} = \lambda_{21}=1$, $\lambda_{12} = \lambda_{22}=0.04$ proposed in~\cite{Measson2004}, we propose the new set : $\lambda_{11} = 1$, $\lambda_{21}=0.2$, $\lambda_{12} = \lambda_{22}=0.07$ and $g$=2, which yields a fit of the same high quality (see fig.~\ref{HT}), but yields also the good values for the gap as deduced from thermal conductivity measurements. Again, it is only the ratio of the $\lambda_{ij}$ which matters, the value $\lambda_{11} =1$ being arbitrarily fixed~\cite{Measson2004}. The factor 5 between $\lambda_{11}$ and $\lambda_{21}$ is essentially due to the coupling strength, meaning that intra band coupling in the band with heavy effective masses (having f character) is much stronger than inter band coupling from this band to the band with a small mass (weak f character). Of course, $\lambda_{12}$ and  $\lambda_{22}$ are strongly reduced by density of states effects, but the general trend which emerges from the new set of $\lambda_{ij}$ imposed by the combination of thermal conductivity~\cite{Seyfarth2007} and $H_{c2}$ results is that multiband effects in PrOs$_{4}$Sb$_{12}$ are coming from the difference in the f character of the bands both through density of states and pairing mechanism effects. This conclusion is quite robust as it relies on measurements independent of the sample homogeneity and the number of transitions.

%valeur de g pour ce fit???

%discuter de publi de ROtundu et idee que m* augmente avec H et effet sur fit.

\section{Conclusion}

Although general characterizations point out to a recurrent double superconducting transition in PrOs$_4$Sb$_{12}$ appearing in the samples with the best $RRR$ and high specific heat jump at the superconducting transition, so in good samples, a study of many samples specially small ones (with a typical size of 100~$\mu$m) shows that its occurrence could be a phenomenon
related to an inhomogeneous effect rather than
to fundamental microscopic mechanisms.
%Indeed three samples exhibit a single sharp superconducting transition at $T_c$\sim$T_{c2}$ with a specific heat jump higher
%than the sum of the two specific ;
  More precisely we think the lowest temperature transition $T_{c2}$ is the intrinsic one.
%Finding the origin of the transition at $T_{c1}$ may be helpful for the future improvement of the quality of the sample.
Based on our 4 circles X-ray diffraction results, we conclude that Pr vacancies are certainly present in the samples and with various percentages which might explain the discrepancy between quantitative specific heat results. But establishing a clear relationship between Pr vacancies and the occurence and magnitude of the jump in specific heat at $T_{c1}$ needs further studies, specially at low temperature.

Finally, the superconducting phase diagram of a single transition sample was determined and fitted with a two-band model. It appears in connection with thermal conductivity results that the multiband effects in PrOs$_{4}$Sb$_{12}$ come from the difference in the f character of the bands both through density of states and pairing mechanism effects.

%Moreover, we could document
%and emphasize a remarkable feature of both%
%the temperature and field derivatives of the C/T
%curves at Tc, namely their Lorentzian form.
%The above results do not explain why doubl
%
%The dependence
%between the splitting of Tc and the grain size (Fig. 1
%of Ref. [ 8 ] ) could also suggest that strains at grain
%or twin boundaries could trigger
%
%clue as

We would like to thank J. P\'{e}caut for preliminary X-rays diffraction study. Ames Laboratory is operated for the U.S. Department of Energy by Iowa State University under Contract No. W-7405-ENG-82. This work was supported by the Director for Energy Research, Office of Basic Energy Sciences. This research was supported by the Grant-in Aid for Scientific Research on the Priority Area "Skutterudites" from MEXT in Japan.

\clearpage
\newpage

\end{document}